\documentclass[prb,twocolumn,showpacs,amsmath,amssymb,superscriptaddress,floatfix]{revtex4}
\usepackage{graphicx}
\usepackage{bm}

\begin{document}
\title{Electrostatic theory for imaging experiments on local
charges in quantum Hall systems}
\author{Ana L. C. Pereira}
\affiliation{Theoretical Physics, University of Oxford, 1 Keble
Road, Oxford OX1 3NP, United Kingdom}
\affiliation{Instituto de
F\'\i sica Gleb Wataghin, Unicamp,  C.P. 6165, 13083-970, Campinas,
Brazil}
\author{J. T. Chalker}
\affiliation{Theoretical Physics, University of Oxford, 1 Keble
Road, Oxford OX1 3NP, United Kingdom}
\date{\today}
\begin{abstract}
We use a simple electrostatic treatment to model recent experiments
on quantum Hall systems, in which charging of localised states
by addition of integer or fractionally-charged quasiparticles is observed.
Treating the localised state as a compressible quantum dot or antidot
embedded in an incompressible background, we calculate the electrostatic
potential in its vicinity as a function of its
charge, and the chemical potential values at which its charge changes.
The results offer a quantitative framework for analysis
of the observations.
\end{abstract}



\maketitle

\section{Introduction}

Recent imaging experiments\cite{ilani,martin} on quantum Hall
systems have resolved individual localised states
and identified discrete charging events in which the
charge of these states jumps when
the mean electron density of the system is altered.
The experiments use a scanning probe carrying a single-electron
transistor\cite{yoo,yacoby_ssc} to measure locally the electrostatic
potential and the compressibility\cite{west} of the two-dimensional
electron gas forming the quantum Hall system.
For a system close to an integer quantum Hall plateau,
the charging events are believed to involve the addition or removal
of a single electron\cite{ilani} to or from the localised state.
Close to a fractional quantum Hall plateau, by contrast,
the observed jumps in localised charge correspond to the
movement of {\it fractionally} charged quasiparticles.\cite{martin}
The latter measurements therefore provide a very direct probe
of these quasiparticles, whose existence is central to the
theory of the fractional quantum Hall effect.\cite{laughlin}

In this paper, motivated by these experiments,
we set out a simple description of a localised
state in a quantum Hall system for a regime where
behaviour is dominated by Coulomb interactions.
We treat interactions using the Thomas-Fermi approximation,
making use of the well-established picture for screening
in integer quantum Hall
systems,\cite{luryi,efros88,deruelle,csg,cms,chalker,burnett,fogler2,fogler3}
in which the sample is divided into compressible
regions where the local Landau-level filling factor
is non-integer, and incompressible regions where the
filling factor is integer. Taking this approach,
a localised variation in charge density, embedded in
an incompressible background, may be induced around a maximum or
minimum in the electrostatic potential due to donors and impurities.
In this way, a quantum dot or antidot is formed with a net charge
that is an integer multiple of the electron charge
for an incompressible background with integer filling
factor. To treat
localised states in fractional quantum Hall systems, we simply assume
that quasiparticle charge replaces electron charge.
While the theory of such quantum dots has been discussed
in some detail previously,\cite{deruelle,burnett,fogler2} and reduces to
a standard problem in electrostatics,\cite{sneddon} a calculation of
quantities relevant for imaging experiments has not, so far as we know,
been presented previously. We hope that the results we describe here
will be useful in further analysis of the observations.

\section{Modelling}

To be definite, we discuss electrons partially
filling a Landau level to form a quantum dot, which has charge density
$\sigma({\bf r})$ as a function
of position $\bf r$ in the plane of the two-dimensional electron gas.
An impurity potential $V_{imp}({\bf r})$ and the screened potential
$V_{scr}({\bf r})$ are related by
\begin{equation}
 V_{scr}({\bf r})=V_{imp}({\bf r}) - \frac{e}{4\pi \varepsilon \varepsilon
_0} \int d^2{\bf r}' \frac{\sigma({\bf r}')}{|{\bf r}-{\bf r}'|}   \,,
 \label{eq1}
\end{equation}
where we denote the electron charge by $-e$.
Using the Thomas-Fermi approximation for a quantum Hall system,
$\sigma({\bf r})=0$ in the incompressible region surrounding the dot.
Throughout the compressible region that makes up the dot, screening
is  perfect and electrons are free to adjust their density so that
$V_{scr}(r)=\mu$, the chemical potential.
The screening charge density is restricted to lie
within the limits $0<-\sigma({\bf r})<\sigma_{max}$,
where $\sigma_{max}$ is the magnitude of the charge density in a filled Landau level;
we consider only impurity potentials flat enough that
this upper limit can be ignored. We choose $V_{imp}({\bf r})$
to have an axially symmetric, parabolic minimum at the origin, so that
\begin{equation}
 V_{imp}(r) = Kr^2
 \label{eq2}
\end{equation}
within the radius $r_d$ of the compressible region.

Imaging experiments\cite{ilani,martin,yacoby_ssc} probe
the  electrostatic potential $\Phi(r,z)$ due to the charge in the
localised state represented by the dot. We idealise the
electron gas as a charge sheet of vanishing thickness, located exactly
at the interface between semiconductor and vacuum, with relative dielectric constants
$\varepsilon_1$ and $\varepsilon_2=1$.
The resulting electrostatic problem is equivalent to one in which there is a single
medium with dielectric constant $\varepsilon =(\varepsilon_1 + \varepsilon_2)$.
This approximation is good provided $r_d$ is large compared with
the thickness of the electron gas and
compared with its depth below the semiconductor surface,
which seems to be the case in the experiments of Ref.~\onlinecite{ilani} and \onlinecite{martin}.
The potential satisfies Laplace's equation
in three dimensions, except on the plane of the electron gas
within the compressible region, where the boundary condition

\begin{equation}
 -e \Phi(r,0) = \mu - V_{imp}(r)   \;\;\;\; \textrm{for} \;\ r \leqslant r_d
 \label{eq3}
\end{equation}

applies. In addition, in the incompressible region, consistency
requires

\begin{equation}
-e \Phi(r,0) > \mu - V_{imp}(r)   \;\;\;\; \textrm{for} \;\ r > r_d
\,.
 \label{eq4}
\end{equation}

The solution can be written in the form\cite{sneddon}

\begin{equation}
\Phi (r,z)= \int_0^\infty dk A(k) J_0(kr)e^{-k|z|} ,
\label{eq6}
\end{equation}

with
\begin{equation}
A (k)= \int_0^{r_d} f(t) \cos(kt) dt.
\label{eq7}
\end{equation}

where, for the parabolic potential of Eq.~(\ref{eq2}),
\begin{equation}
f(t) =  \frac{2}{\pi e} (2 K t^2 - \mu). \label{eq8}
\end{equation}

The charge density in the compressible region is determined from
\begin{equation}
\frac{\sigma(r)}{\varepsilon \varepsilon_0}= -
\frac{d\Phi}{dz}\bigg{|}_{z=0^+} +  \frac{d\Phi}{dz}\bigg{|}_{z=0^-}
\label{eq9}
\end{equation}
for $r \leq r_d$.
The value of the chemical potential is fixed by the requirements
that Eq.~(\ref{eq4}) is satisfied and that there is no divergence in
in the charge density: it is
\begin{equation}
\mu=2K r_d^2.
\label{mi}
\end{equation}
With this, the charge density for $r \leqslant r_d$
is
\begin{equation}
\sigma(r)= - \frac{16 K \varepsilon\varepsilon_0}{\pi e}
\sqrt{r_d^2 -  r^2} 
\end{equation}
(a result given previously in, for example, Ref.~\onlinecite{fogler2})
and the total charge on the dot is
\begin{equation}
Q = 2\pi \int_0^{r_d} r \, dr \, \sigma(r)=-e\left(\frac{r_d}{\alpha}\right)^3\,,
\label{Q}
\end{equation}
where we have introduced the length scale
\begin{equation}
\alpha \equiv {\Big(}\frac{3e^2}{32K \varepsilon \varepsilon_0}{\Big
)}^{1/3}\,. \label{alpha}
\end{equation}

It is also useful to calculate the total energy $E(Q)$ of the charge on the dot,
which can be done by integrating the relation $\mu = -e \partial E(Q)/\partial Q$,
using Eqns.~(\ref{mi}) and (\ref{Q}). We find
\begin{equation}
E(Q)=\frac{6}{5}K\alpha^2\left(\frac{-Q}{e}\right)^{5/3}\,.
\label{energy}
\end{equation}

At this stage, we take account of the fact that charge is discrete by setting
$Q=-Ne$, where the number of
electrons contained in the dot is $N=1,2,3,...$
In consequence, the dot radius
takes the values
\begin{equation}
r_d = \alpha {N}^{1/3} . \label{radius}
\end{equation}
Having restricted the charge to these discrete values, $\Phi(r,0)$
is no longer related to the chemical potential for the sample
by Eq.~(\ref{eq3}): instead, combining Eq.~(\ref{eq3}) and Eq.~(\ref{mi}), one has
\begin{equation}
 -e \Phi(r,0) = 2Kr_d^2 - V_{imp}(r)   \;\;\;\; \textrm{for} \;\ r \leqslant r_d \,.
\nonumber
\end{equation}
The values of $\mu$ at which charge jumps occur can be found by
minimising the free energy $F=E(Q) - \mu N$ of the dot in equilibrium
with a charge reservoir, over integer $N$, and considering the result as a function of
$\mu$. From the expression
\begin{equation}
F=\frac{6}{5}K\alpha^2N^{5/3} -\mu N \label{free_energy}
\end{equation}
we find that
the values of $\mu$ at which the occupation of the dot changes between
$N$ and $N+1$ are
\begin{equation}
\mu_{N \leftrightarrow N+1}=\frac{6}{5} K\alpha^2
[(N+1)^{5/3}-N^{5/3}]\,. \label{michange}
\end{equation}

Next we evaluate the electrostatic potential $\Phi(r,z)$.
Combining Eqns.~(\ref{eq6}), (\ref{eq8}) and (\ref{radius}), we have
\begin{eqnarray}
\Phi_N(r,z) & = & -  \frac{4 K}{\pi e}
\int_0^{\alpha \sqrt[3]{N}} dt (\alpha^2 N^{2/3} - t^2)\times   \nonumber \\
 & & \times \int_0^\infty e^{-k|z|} \cos (kt) J_0(kr) dk \, .
\label{phirz}
\end{eqnarray}
In this expression, the integral on $k$ can be evaluated analytically
but the one on $t$ must be done numerically.
The result can be written in terms of the scaled variables
$\rho=r/\alpha$, $\zeta=z/\alpha$ and $\tau=t/\alpha$ as

\begin{equation}
\Phi_N(\alpha\rho,\alpha\zeta)=  \frac{-e}{4 \pi \varepsilon
\varepsilon_0 \alpha}F_N(\rho,\zeta)
\label{phi}
\end{equation}

with
\begin{eqnarray}
F_N(\rho,\zeta) = \frac{3}{2} \int_0^{ \sqrt[3]{N}} \!\!\!\!\! d\tau
(N^{2/3}-\tau^2)
\Bigg[\frac{\sqrt{\lambda^4+4\tau^2\zeta^2}+\lambda^2}{2\lambda^4+8\tau^2\zeta^2)}
\Bigg]^{1/2} \nonumber
\end{eqnarray}
where $\lambda^2 = \zeta^2+\rho^2-\tau^2$.

Far from the dot,
for $(\rho^2 +\zeta^2) >> N^{2/3}$, these expressions reduce
to
\begin{equation}
\Phi_N(r,z) = \frac{-Ne}{4 \pi \varepsilon \varepsilon_0 } \frac{1}{
\sqrt{r^2+z^2}} \; ,
\end{equation}
as expected.

\begin{figure}[htb]
\includegraphics[width=8.6cm, height=6.3cm]{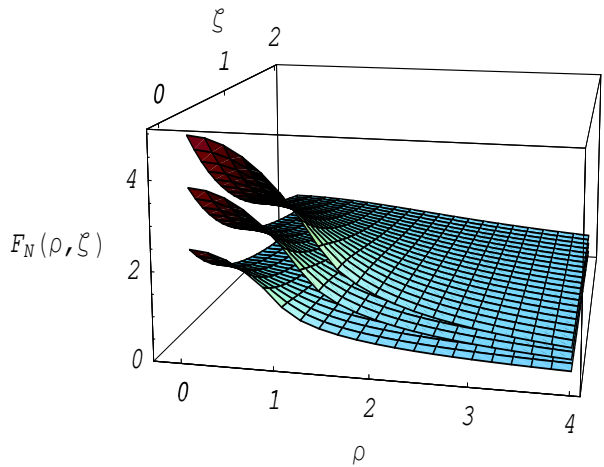}
\caption{The scaled electrostatic potential $F_N(\rho,\zeta)$
as a function of scaled radius $\rho$  and height $\zeta$ from the centre of the dot,
for $N=1$ (lower surface), $N=2$ and $N=3$ (upper
surface).} \label{fig2}
\end{figure}

\begin{figure}[htb]
\includegraphics[width=6.5cm,height=4.6cm]{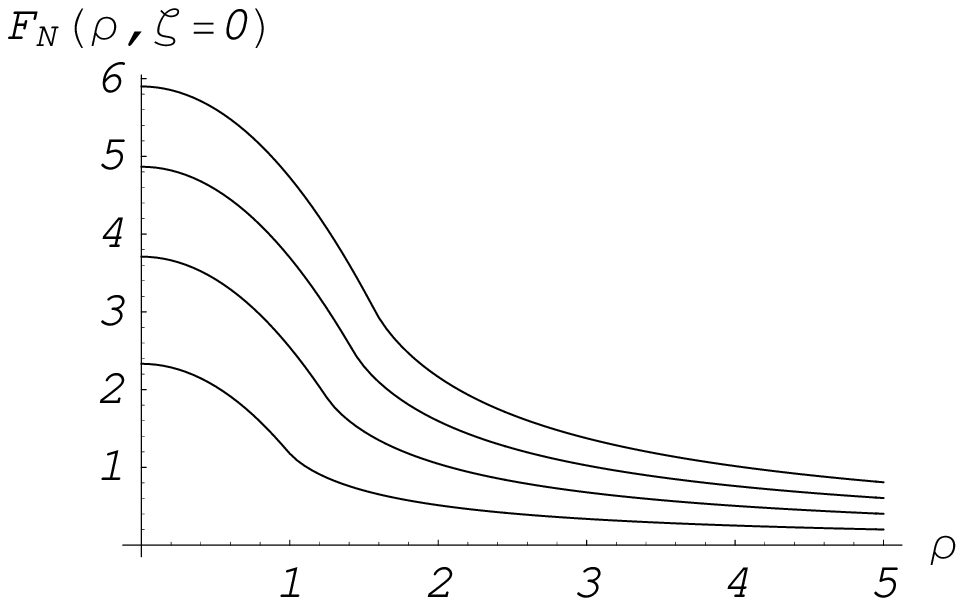}
\includegraphics[width=6.5cm,height=4.6cm]{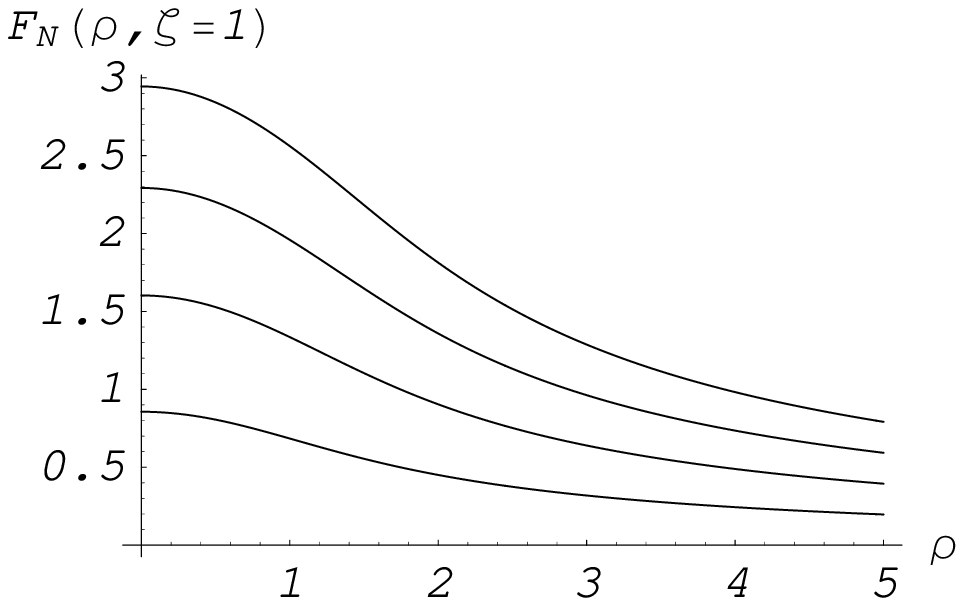}
\caption{$F_N$ as a function of $\rho$, at fixed height $\zeta$ above
the dot. In each graph, the curves are for $N=1$ (lowest),
$N=2$, $N=3$, and $N=4$ (highest).} \label{fig3}
\end{figure}

The dependence of the function $F_N(\rho,\zeta)$ on $\rho$ and $\zeta$
is illustrated in Fig.~\ref{fig2}, and its variation with $\rho$ at
fixed $\zeta$ is shown in Fig.~\ref{fig3}.

\begin{figure}[htb]
\includegraphics[width=6.5cm,height=4.6cm]{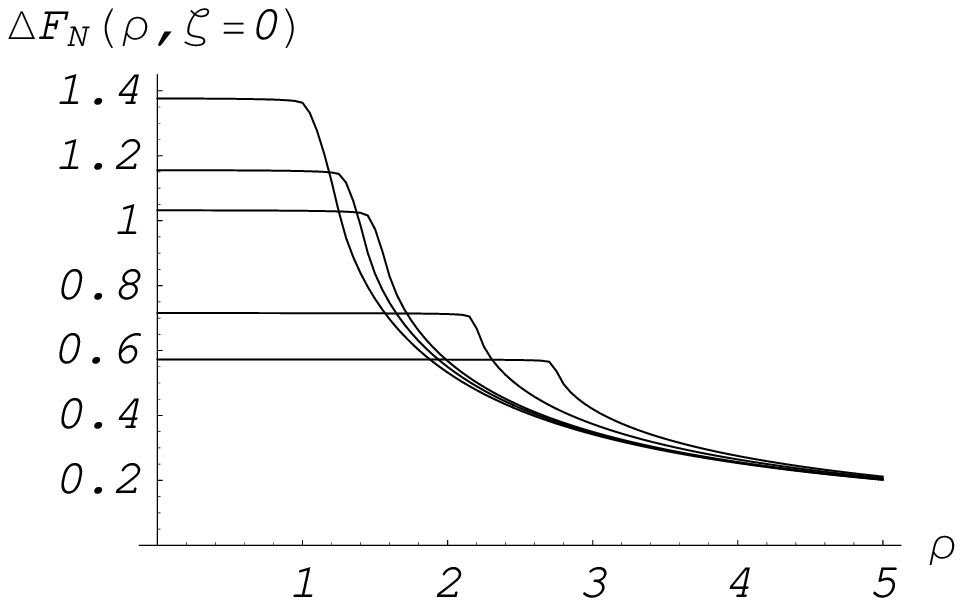}
\includegraphics[width=6.5cm,height=4.6cm]{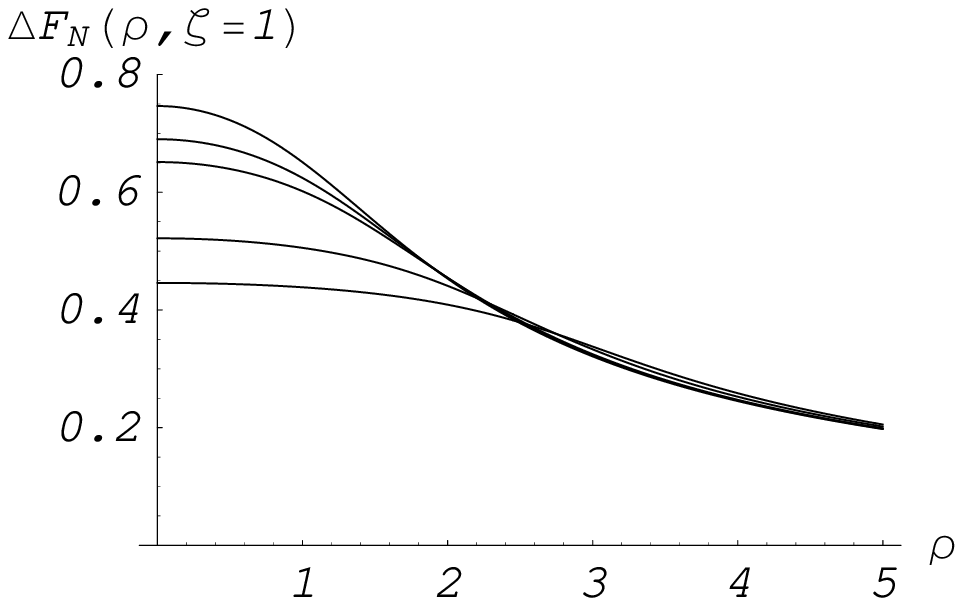}
\caption{Electrostatic potential changes (proportional to
$\Delta F_N$) produced by jumps in the charge of the dot, as a function
of $\rho$, at fixed distances, $\zeta = 0$ and $\zeta = 1$ above the dot.
In each graph,
the different curves are for $N=1$ (the highest curve at
$\rho=0$), $N=2$, $N=3$, $N=10$ and $N=20$
(the lowest curve at $\rho=0$) .} \label{fig5}
\end{figure}

Since in experiment this potential will add to other contributions,
for example, from fixed background charges, it is useful
to focus on the potential changes arising from jumps in
the charge of the dot. These changes are
proportional to $\Delta F_N(\rho,\zeta) \equiv F_{N+1}(\rho,\zeta) - F_{N}(\rho,\zeta)$,
and this function is shown in Fig.~\ref{fig5}.

\section{Discussion}
There is scope to compare these results with experiment in several different ways.

First, the most striking feature of the observations is the fact that,
considering behaviour as a function of average electron density $n$ and
flux density $B$, a particular charging event takes place on a line
in the $n$-$B$ plane which is parallel to one of the lines
of integer filling factor $\nu$. Such behaviour is built into the
model we have studied. In particular, suppose that
$N_L$ Landau levels in the sample are completely filled, so that
the charge density  within the quantum dot ($\sigma({\bf r})$ in Eq.~(\ref{eq1}))
lies in the $(N_L+1)$ th level. In that case, if $n$ and $B$ vary together
along a line in the $n$-$B$ plane parallel to $\nu = N_L$, the charge density
variation in the sample is uniform in space and the screened potential remains constant;
as a consequence, the charge of the quantum dot is unchanged.
Conversely, charging events are produced by moving in a perpendicular
direction in the $n$-$B$ plane. This account omits
the single-particle contribution, $(N_L+1/2)\hbar \omega_c$, to the energy of
the charge within the dot. The approximation is justified because the electrostatic
part of the energy of the two-dimensional electron gas is dominant.

Turning to more specific aspects of our modelling, it is useful to
focus on results that are independent of the model parameter $K$ in Eq.~(\ref{eq2})
and of measurement calibration. Two such results (which are physically related
to each other) are the
power laws appearing in the dependence of dot radius
on electron number, Eq.~(\ref{radius}), and in the chemical potential values
at which charge jumps occur, Eq.~(\ref{michange}).
While dot radius is probably difficult to measure precisely, because of
issues of resolution, as discussed below, the relative size of chemical potential steps
required to add a sequence of charges should be an accessible quantity.
Specifically, charge jumps are produced experimentally
by a change in the backgate voltage applied to a sample:\cite{ilani,martin}
fitting the ratio of voltage steps for successive jumps to
Eq.~(\ref{michange}) would provide a test of the theory we have presented and
a determination of the number of electrons within the dot.
Deviations from the theory would arise either if the confining
potential is not parabolic, or, more interestingly, if
many-body correlations within the compressible region, which are
omitted from Thomas-Fermi theory, make an important contribution to the
total energy of the electrons in the dot. Even in these cases, we expect
as a robust feature a decrease in the size of voltage steps between charge jumps
as electron number increases.

In addition, one can attempt an absolute comparison of
theoretical and experimental quantities. As an illustration,
suppose $\alpha=200$ nm and $N=10$, so that $r_d=430$ nm,
and consider a measurement of the potential by a scanning
probe at a height $z=200$ nm above the sample.
Taking, for GaAs, $\varepsilon_1=13$, we find from Eq.~(\ref{phi})
a change in electrostatic potential when a further electron is added,
of size
$\Delta\Phi_{10}(r=0,z=200 {\rm nm})=520 \mu {\rm V}$.
This is similar to the step size of $180 \mu {\rm V}$ reported
in Ref.~\onlinecite{martin}; an exact match could presumably
be arranged by adjusting $\alpha$, $N$ or $z$.
Beyond this, one can regard our calculation of $\Delta F(\rho,\zeta)$,
illustrated in Fig.~\ref{fig5}, as a determination of the
resolution function for the imaging technique.

In summary, we have presented a simple model for the
imaging experiments of Ref. \onlinecite{ilani}
and \onlinecite{martin}. A closer comparison between
observations and calculations should help
determine the numbers of electrons contained in localised states
and the spatial size of these states, while deviations of measurements
from this theory may be an indication of correlation effects.

We are grateful to Amir Yacoby for discussions and
for preprints of Refs.~\onlinecite{ilani}
and \onlinecite{martin}. The work was supported in part by CAPES,
and by EPSRC under Grant GR/R83712/01.

\end{document}